\documentclass[aps,twocolumn,prl,showpacs,superscriptaddress,groupedaddress]{revtex4}  
\usepackage{dcolumn}
\usepackage{bm}
\usepackage{color}

\def\ltape{\hbox{\ $<$\hskip -8pt\raise -4pt\hbox{$\sim$}\ }}
\def\gtape{\hbox{\ $>$\hskip -8pt\raise -4pt\hbox{$\sim$}\ }}

   \ifx\pdfoutput\undefined
   \usepackage[dvips]{graphicx}
   \else
   \usepackage[pdftex]{graphicx}
   \fi
   \usepackage{epstopdf}
   \DeclareGraphicsExtensions{.pdf,.gif,.jpg,.eps,.png}
  
   \usepackage{amsmath}
  \usepackage{amssymb}

\pdfoutput=1

\begin{document}

\title{3D Magnetic Reconnection with a spatially confined X-line extent -- Implications for Dipolarizing Flux Bundles and the Dawn-Dusk Asymmetry}

\author{Yi-Hsin~Liu}
\affiliation{Dartmouth College, Hanover, NH 03750}
\author{Tak Chu~Li}
\affiliation{Dartmouth College, Hanover, NH 03750}
\author{M.~Hesse}
\affiliation{University of Bergen, Bergen, Norway}
\affiliation{Southwest Research Institute, San Antonio, TX 78238}
\author{Weijie~Sun}
\affiliation{University of Michigan, Ann Arbor, MI 48109}
\author{Jiang~Liu}
\affiliation{University of California, Los Angeles, CA 90095}
\author{J.~Burch}
\affiliation{Southwest Research Institute, San Antonio, TX 78238}
\author{J. A. ~Slavin}
\affiliation{University of Michigan, Ann Arbor, MI 48109}
\author{Kai~Huang}
\affiliation{Dartmouth College, Hanover, NH 03750}
\affiliation{University of Science and Technology of China, China}

\begin{abstract}

Using 3D particle-in-cell (PIC) simulations, we study magnetic reconnection with the x-line being spatially confined in the current direction. We include thick current layers to prevent reconnection at two ends of a thin current sheet that has a thickness on an ion inertial ($d_i$) scale. The reconnection rate and outflow speed drop significantly when the extent of the thin current sheet in the current direction is $\lesssim O(10 d_i)$. When the thin current sheet extent is long enough, we find it consists of two distinct regions; an inactive region (on the ion-drifting side) exists adjacent to the active region where reconnection proceeds normally as in a 2D case. The extent of this inactive region is $\simeq O(10 d_i)$, and it suppresses reconnection when the thin current sheet extent is comparable or shorter. The time-scale of current sheet thinning toward fast reconnection can be translated into the spatial-scale of this inactive region; because electron drifts inside the ion diffusion region transport the reconnected magnetic flux, that drives outflows and furthers the current sheet thinning, away from this region. This is a consequence of the Hall effect in 3D. While this inactive region may explain the shortest possible azimuthal extent of dipolarizing flux bundles at Earth, it may also explain the dawn-dusk asymmetry observed at the magnetotail of Mercury, that has a global dawn-dusk extent much shorter than that of Earth. 

\end{abstract}

\pacs{52.27.Ny, 52.35.Vd, 98.54.Cm, 98.70.Rz}

\maketitle

{ \it \bf 1. Introduction--}
Through changing the magnetic connectivity, magnetic reconnection converts magnetic energy into plasma kinetic and thermal energies. It plays a critical role in the energy release of geomagnetic substorms both in Earth \cite{baker96a,angelopoulos08a} and other planets \cite{slavin10a,WSun15a, kronberg05a, mitchell05a,southwood16a}. During reconnection, the magnetic connectivity is altered at geometrically special points, that constitute a ``reconnection x-line'' in the current direction. In a two-dimensional (2D) model, the extent of the reconnection x-line is, technically, {\it infinitely} long due to the translational invariance out of the reconnection plane. It is of great interests to understand the fundamental nature of a three-dimensional (3D) reconnection in the opposite limit. Especially, it remains unclear how a spatial confinement in the current direction would affect reconnection and whether there is a minimal requirement for the spatial extent of the reconnection x-line. Spatially confined reconnection can be relevant to azimuthally localized dipolarizing flux bundles (DFBs) at Earth's magnetotail \cite{JLiu13b,SLi14a} and Mercury's entire magnetotail that has a short dawn-dusk extent \cite{WSun16a, poh17b, rong18a}.

DFBs are magnetic flux tubes embedded in fast earthward flows called bursty bulk flows (BBFs), and the leading edge of each DFB has been termed a dipolarization front (DF). Observations show that they are localized in the azimuthal (i.e., dawn-dusk) direction with a typical extent of  $3 R_E$ \cite{nakamura04a,nagai13a,JLiu13b,SLi14a}, and the shortest extent observed is $\simeq 0.5R_E \simeq 10 d_i$ \cite{JLiu15b}. Here $R_E$ is the Earth's radius and $d_i$ is the ion inertial length. These fast earthward flows are observed during substorms and have been associated observationally with Pi2 pulsations and the substorm current wedge (e.g., \cite{kepko15a} and references therein). A localized DFB could originate from (1) an initially long dawn-dusk extended DFB that breaks up into smaller pieces (through interchange/ballooning instability) during the intrusion into the inner tail \cite{birn11a,lapenta11a,pritchett14a,sitnov14a,birn15a}, or, (2) simply from an azimuthally localized reconnection x-line, where the {\it frozen-in} condition is violated \cite{shay03a,pritchett13a,pritchett18a} within a finite azimuthal extent. While both mechanisms are possible in nature, in this work we study scenario (2) using a simple setup. In addition, spatially confined reconnection also has a direct application to the magnetotails of other planets, such as Mercury, whose global dawn-dusk extent is as short as a few 10's of $d_i$ \cite{WSun16a, poh17b, rong18a}. Interestingly, observations by MESSENGER \cite{WSun16a} indicate a higher occurrence rate of DFBs on the dawn side of Mercury's magnetotail, opposite to that observed at Earth's magnetotail (whose extent is a few 100's of $d_i$). An explanation to this dawn-dusk asymmetry is desirable.

Previous attempts that model the effect of the dawn-dusk localization on reconnection and bursty bulk flows are briefly summarized here. Shay et al. \cite{shay03a} used initial perturbation spatially localized in the current direction to induce reconnection in two-fluid simulations. The shortest reconnection x-line in their simulation is $\simeq 10d_i$ long, but the x-line spreads in the current direction unless the initial uniform current sheet is thicker than $4d_i$. In a follow-up study, Meyer et al. \cite{meyer15a} derived a model of the outflow speed reduction using Sweet-Parker type analysis in 3D diffusion regions. Dorfman et al. \cite{dorfman14a} studied the localized reconnection region experimentally in MRX. More recently, Arnold et al. \cite{arnold18a} used a 2D Riemann setup to study the outflow reduction and suggested that the ion momentum transfer from the ion drifting direction to the outflow direction is critical. Pritchett and Lu \cite{pritchett18a} used a localized driving to study reconnection onset in tail geometry.

To study the effect of the dawn-dusk localization on reconnection, we confine the reconnection region by embedding a thin reconnecting current sheet between much thicker sheets. This spatial confinement strongly limits the spread of the x-line \cite{shay03a,TKMNakamura12a,shepherd12a,TCLi19a}. This machinery allows us to study the 3D nature of reconnection as a function of the x-line extent in a controlled fashion. Our simulations demonstrate that reconnection is strongly suppressed if the thin current sheet extent is shorter than a critical length of $\simeq O(10 d_i)$. Through detailed examinations of thin reconnecting current sheets of extent $31 d_i$ and $8.4d_i$, we link this critical confinement scale to the extent of an inactive region on the ion-drifting side of the x-line, that connects to an active reconnection region on the electron-drift side. This two-region structure develops because the reconnected magnetic flux, that drives outflows and furthers the current sheet thinning, is preferentially transported by electrons in the direction of the electron drift. We show that the time-scale toward fast reconnection can be translated into the spatial-scale of this inactive region. This shortest possible x-line extent of $\sim O(10 d_i)$ for fast reconnection manifested here can be relevant to the narrowest BBFs/DFBs observed at Earth's magnetotail \cite{JLiu15b}. In addition, since the dawn-dusk extent of the entire magnetotail of Mercury is similar to the case considered here, the preferential transport of the reconnected magnetic flux to the electron-drifting side (i.e., the dawn side) can explain the observed dawn-dusk asymmetry of the occurrence rate of DFBs \cite{WSun16a,WSun17a}. In the end, we incorporate the dawn-dusk asymmetry argument in Lu et al. \cite{SLu16a,SLu18a}, and propose that the opposite dawn-dusk asymmetry observed at Mercury and Earth is primarily caused by the vastly different global dawn-dusk scale.

The structure of this paper is outlined in the following. Section 2 describes the simulation setup. Section 3 shows the scaling of reconnection as a function of the confinement length scale. Section 4 shows the details of a case with a long confinement scale. Section 5 shows the details of a case with a short confinement scale. In section 6, we pin down the underlying physics that determines the critical confinement scale for suppression; section 6.1 examines the 3D generalized Ohm's law. Section 6.2 examines the flux transport and the asymmetric thinning. Section 7 summarizes our results and proposes our explanation of the dawn-dusk asymmetry in planetary magnetotails. \\

\begin{figure}
\includegraphics[width=8.0cm]{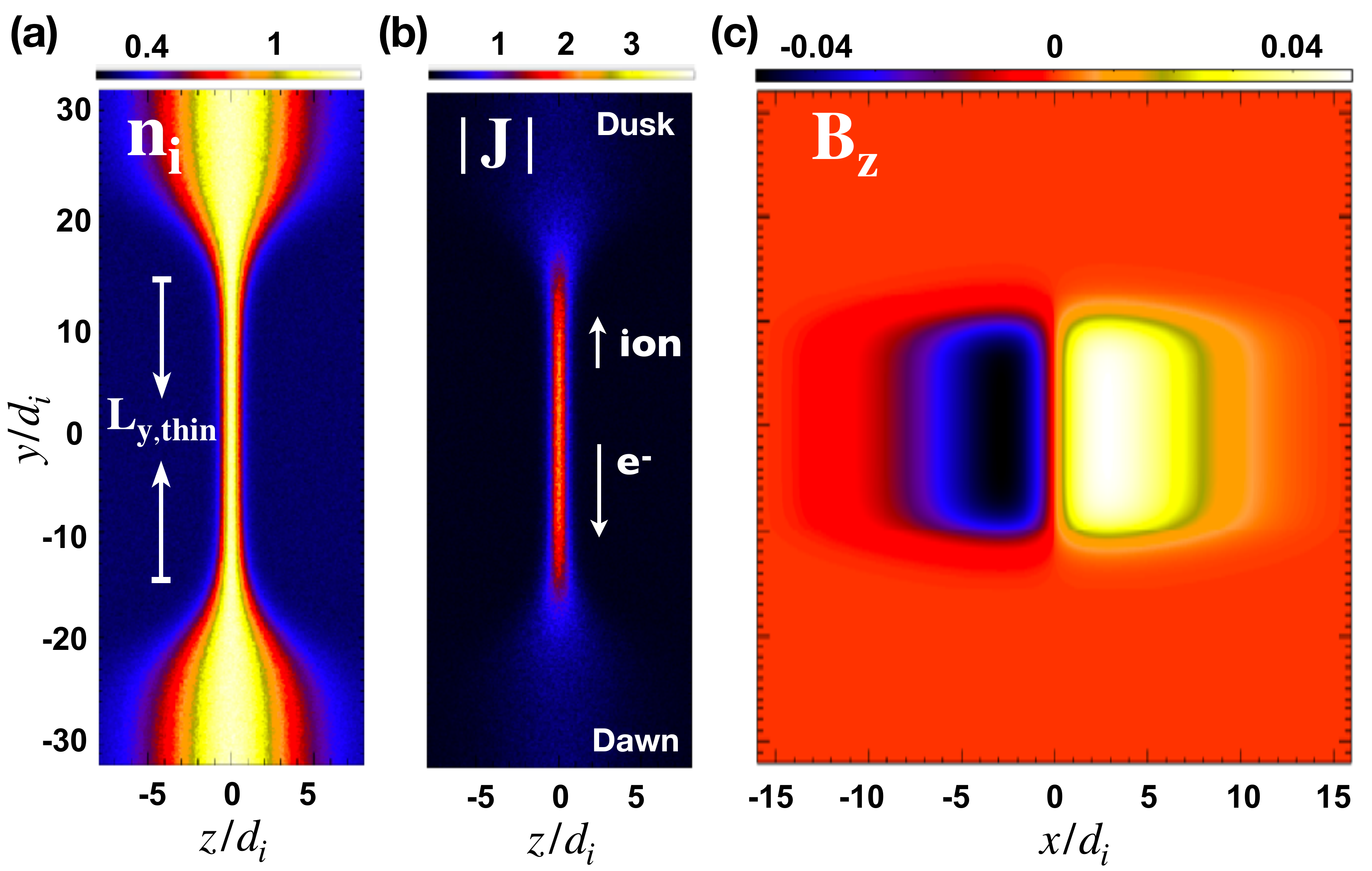} 
\caption {An example setup with $L_{y,thin}=31d_i$. The ion density $n_i$ in (a). The total current $|{\bf J}|$ in (b). The initial magnetic perturbation $B_z$ in (c).
}
\label{initial}
\end{figure}

{\it \bf 2. Simulation setup--}
The initial condition consists of the magnetic field ${\bf B}(y,z)=B_0\mbox{tanh}[z/L(y)]\hat{{\bf x}}$ and the plasma density $n(y,z)=n_0\mbox{sech}[z/L(y)]+n_b$. Here the sheet half-thickness $L(y)$ $=L_{min}+(L_{max}-L_{min})[1-f(y)]$ with the function $f(y)=[\mbox{tanh}((y+w_0)/S)-\mbox{tanh}((y-w_0)/S)]/[2\mbox{tanh}(w_0/S)]$. We choose $L_{min}=0.5d_i$, $L_{max}=4d_i$ and $S=5d_i$ and the background density $n_b=0.3n_0$, which will embed a thin sheet of thickness $1d_i(= 2\times L_{min})$ between the ambient thicker sheets of thickness $8d_i (= 2 \times L_{max})$ in the y-direction. In this work, we conduct runs with $w_0=20d_i,10d_i,7.5d_i$ and $2d_i$. We define the length of the thin current sheet $L_{y,thin}$ as the region for $L < 2 \times L_{min}=1d_i$, then the corresponding $L_{y,thin}=31d_i,12d_i, 8.4d_i$ and $4d_i$. We will use $L_{y,thin}$ to label the four runs discussed in this paper. For instance, the initial profiles of the $L_{y,thin}=31d_i$ case is shown in Fig.~\ref{initial} for illustration. Fig.~\ref{initial}(a) shows the density profile with $L_{y,thin}$ marked. Fig.~\ref{initial}(b) shows the total current density $|{\bf J}|$. Note that ions (electrons) drift in the positive (negative) y-direction that corresponds to the dusk (dawn) side at Earth's magnetotail. In addition to the y-varying current sheet thickness, we initiate reconnection with an initial perturbation within the thin current sheet region, as shown in Fig.~\ref{initial}(c). These four simulations have the domain size $L_x \times L_y \times L_z = 32d_i \times 64d_i \times 16d_i$ and $768 \times 1536 \times 384$ cells. The mass ratio is $m_i/m_e = 75$. The ratio of the electron plasma to gyrofrequency is $\omega_{pe} /\Omega_{ce} = 4$ where $\omega_{pe} \equiv (4\pi n_0 e^2/m_e)^{1/2}$ and $\Omega_{ce}\equiv eB_0/m_e c$. In the presentation, densities, time, velocities, spatial scales, magnetic fields, and electric fields are normalized to $n_0$, the ion gyrofrequency $\Omega_{ci}$, the Alfv\'enic speed $V_A=B_0/(4\pi n_0 m_i)^{1/2}$, the ion inertia length $d_i=c/\omega_{pi}$, the reconnecting field $B_0$ and $V_AB_0/c$, respectively. The boundary conditions are periodic both in the x- and y-directions, while in the z-direction they are conducting for fields and reflecting for particles.

This setup will confine magnetic reconnection within the thin sheet region and prevent the reconnection x-line from progressively spreading into two ends [e.g., \cite{TCLi19a}]. Plasmas are loaded as drifting Maxwellians that satisfy the total pressure $P+B^2/8\pi=const$, and drifting speeds satisfy ${\bf J}=en({\bf V}_{id}-{\bf V}_{ed})=(c/4\pi) \nabla\times {\bf B}$ and $V_{id}/V_{ed}=T_i/T_e$ as in the standard Harris sheet equilibrium \cite{harris62}. These satisfy the relation ${\bf J}\times {\bf B}+\nabla P=0$ and  $\nabla\cdot {\bf B}=0$. Note that the inertial force $m_i V_{iy}\partial_yV_{iy}$ in the transition regions (i.e., where $L(y)$ varies) does not vanish. To reduce this force that could move the entire structure in the +y-direction, we load an uniform ion drift velocity $V_{iy}$ with a value that satisfies the Harris equilibrium at the ambient thicker sheet that has $L=L_{max}=4d_i$ and $(T_i/T_e)_{thick}=5$. This setup gets closer to an equilibrium in the limit of small $V_{iy}$ that can be satisfied when the ambient thicker sheet is thick enough. A small drifting speed $V_{iy}$ also reduces the drift-kink instability arising from ion shear flows between the ambient and sheet regions \cite{karimabadi03a}. One may expect that particles would just stream out of the thin current sheet region, making the setup fall apart. However, it is not this case since the primary current carrier drift is the diamagnetic drift, where the guiding centers do not move. \\

\begin{figure}
\includegraphics[width=7.0cm]{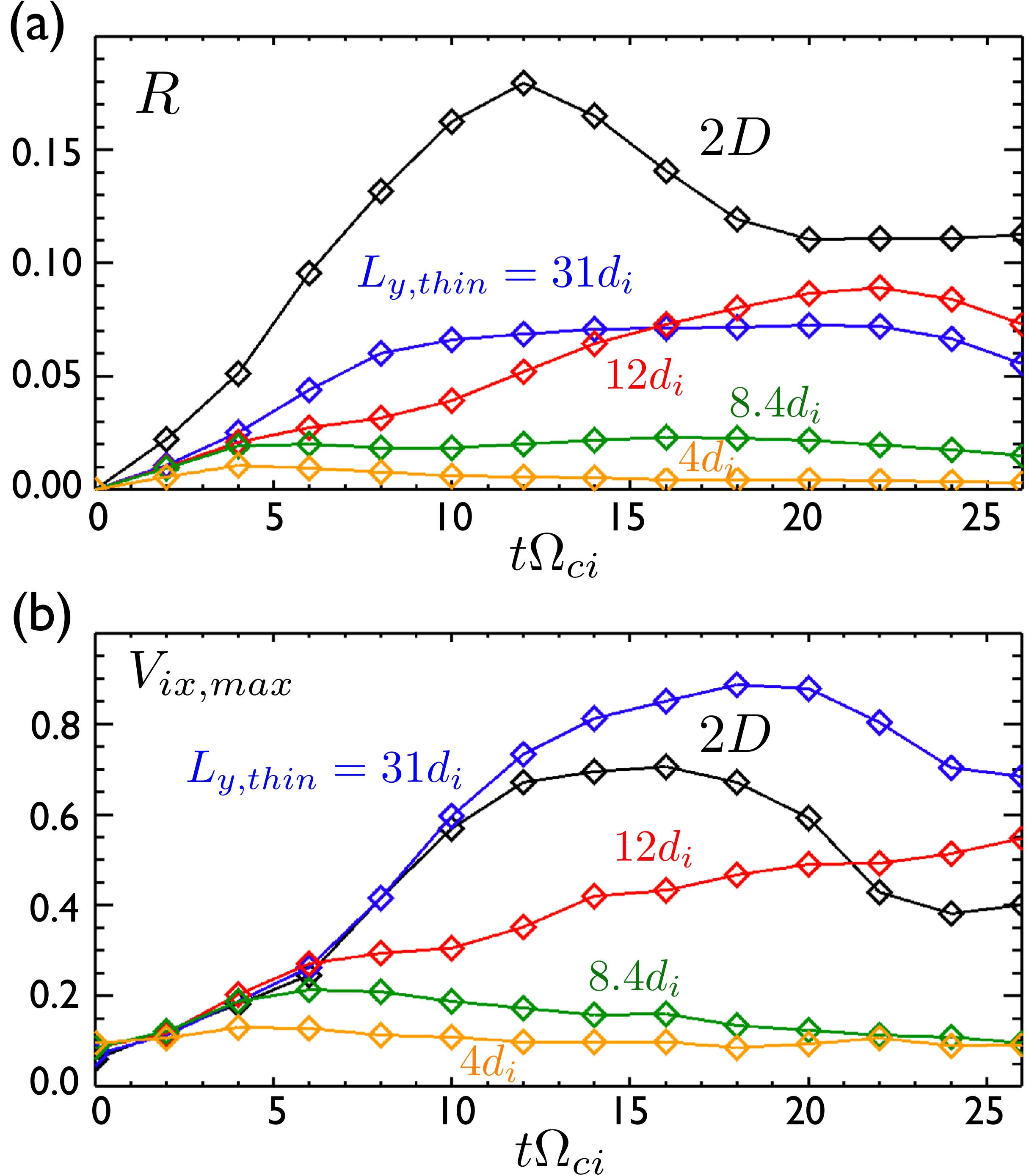} 
\caption {The time evolution of the normalized reconnection rate $R$ and the maximum ion outflow speed $V_{ix,max}$ with different confinement scale $L_{y,thin}$.}
\label{rates}
\end{figure}

{\it \bf 3. Scaling of reconnection rates and outflow speeds--}
With this simulation setup, we can explore how reconnection rates and reconnection outflow speeds are affected by the confinement in the current direction. The results with $L_{y,thin}=31d_i, 12d_i, 8.4d_i$ and $4d_i$ are shown in Fig.~\ref{rates}. For comparison, the companion 2D case is also plotted in black. Given the symmetry of the system in the inflow direction, we can measure the reconnection rate using the increasing rate of the reconnected flux at the $z=0$ plane; the total reconnected flux is $\Psi=\int_{0}^{L_x/2}\int_{-L_y/2}^{L_y/2} B_z(z=0) dxdy$, then the increasing rate of the reconnected flux is $d\Psi/dt$. To compare with 2D, we define the reconnection rate as $R\equiv (d\Psi/dt)/L_{y,thin}$. For the $L_{y,thin}=31d_i$ and $12d_i$ cases, both the reconnection rate and the maximum outflow speed are comparable to that in 2D, where the x-line extent is infinitely long. For the $L_{y,thin}=8.4d_i$ and $4d_i$ cases, we observe the significant impact from the reconnection region confinement, where both the rate and outflow speed plunge into much lower values. These suggest that the critical confinement scale that suppresses reconnection is $\lesssim 10d_i$. In the following, we look into the details of how reconnection works in two cases. The $L_{y,thin}=31d_i$ case has realized 2D-like fast reconnection in part of the thin current sheet, while the $L_{y,thin}=8.4d_i$ case shows reconnection being strongly suppressed. \\

\begin{figure}
\includegraphics[width=8.0cm]{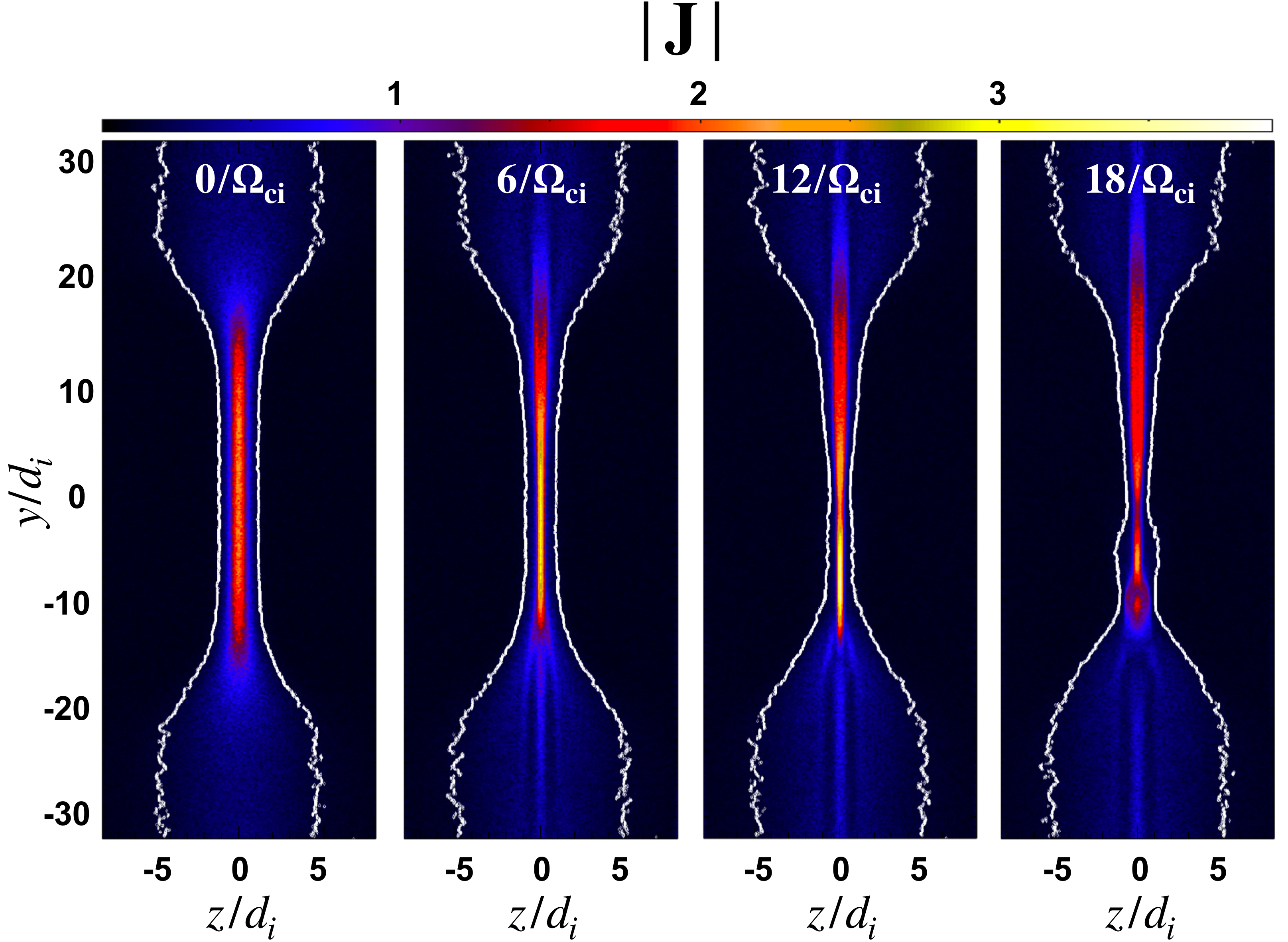} 
\caption {The evolution of the current density $|{\bf J}|$ at the $x=0$ plane inside the 3D box with $L_{y,thin}$$=31d_i$. The white curves trace the boundary of the current sheet.
}
\label{L20_evolve}
\end{figure}

\begin{figure}
\includegraphics[width=8.5cm]{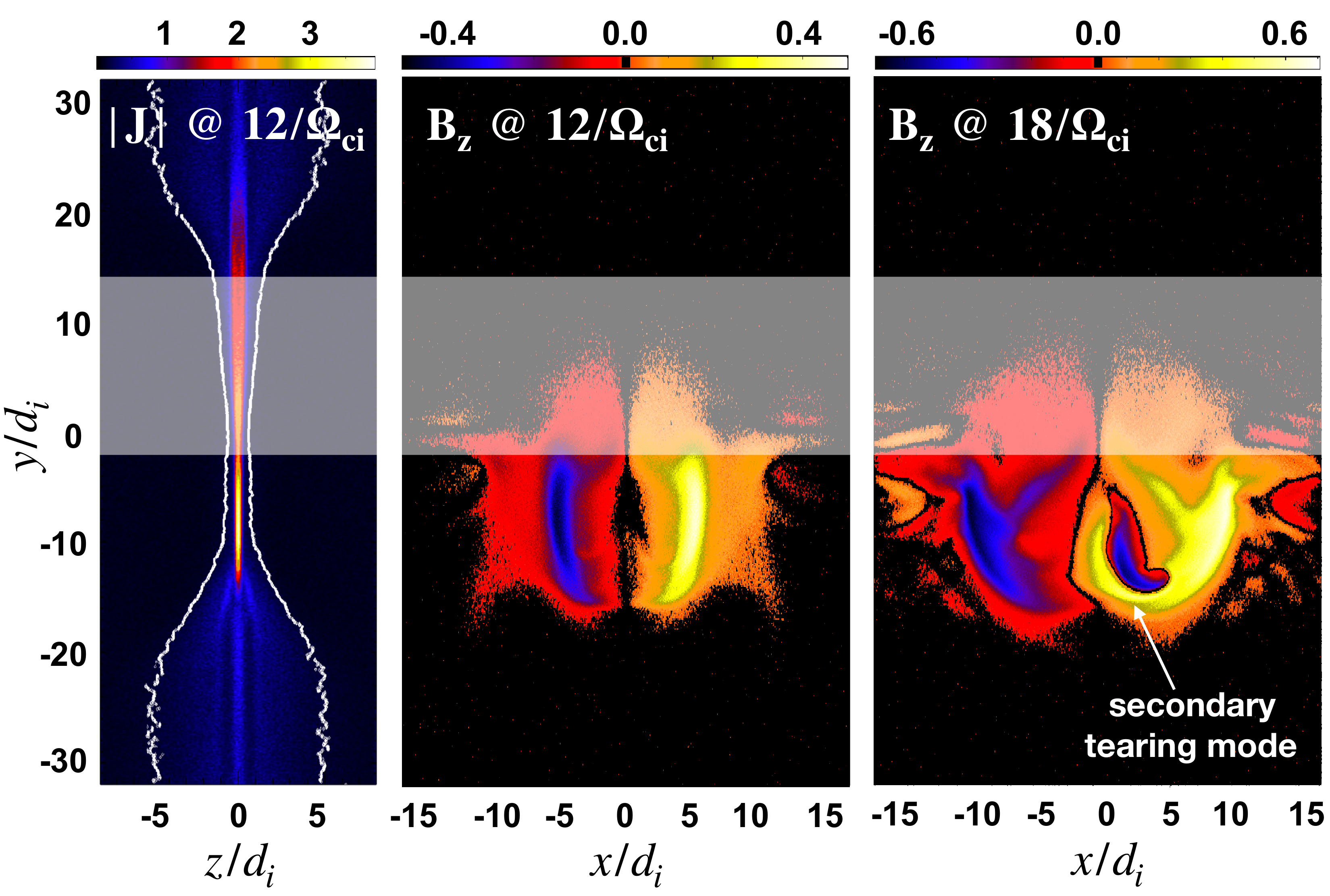} 
\caption {The structure of the reconnected magnetic field $B_z$ at the $x=0$ plane at later time in the $L_{y,thin}$$=31d_i$ case.
}
\label{tearing}
\end{figure}

{\it \bf 4. $L_{y,thin}=31d_i$ case--}
We show the evolution of the total current density $|\bf{J}|$ of the $L_{y,thin}=31 d_i$ case at the $x=0$ plane (right through the x-line) in Fig.~\ref{L20_evolve}. The corresponding times are $0/\Omega_{ci}$, $6/\Omega_{ci}$, $12/\Omega_{ci}$, and $18/\Omega_{ci}$. The boundary of the current sheet, where $J_y$ is slightly larger than the background noise level, is marked by the white curves.
Note that for $z-y$ slice plots throughout this manuscript, ions are drifting upwardly (in the positive y-direction) while electrons are drifting downwardly (in the negative y-direction). We use the same color range for all plots of $|J|$ to facilitate the comparison of the current sheet thinning process. The current sheet thins asymmetrically and leads to a thinner sheet on the electron-drifting side. The bulge at time $18/\Omega_{ci}$ is caused by the generation of a secondary tearing mode, that will be discussed later in Fig.~\ref{tearing}.

\begin{figure}
\includegraphics[width=9.0cm]{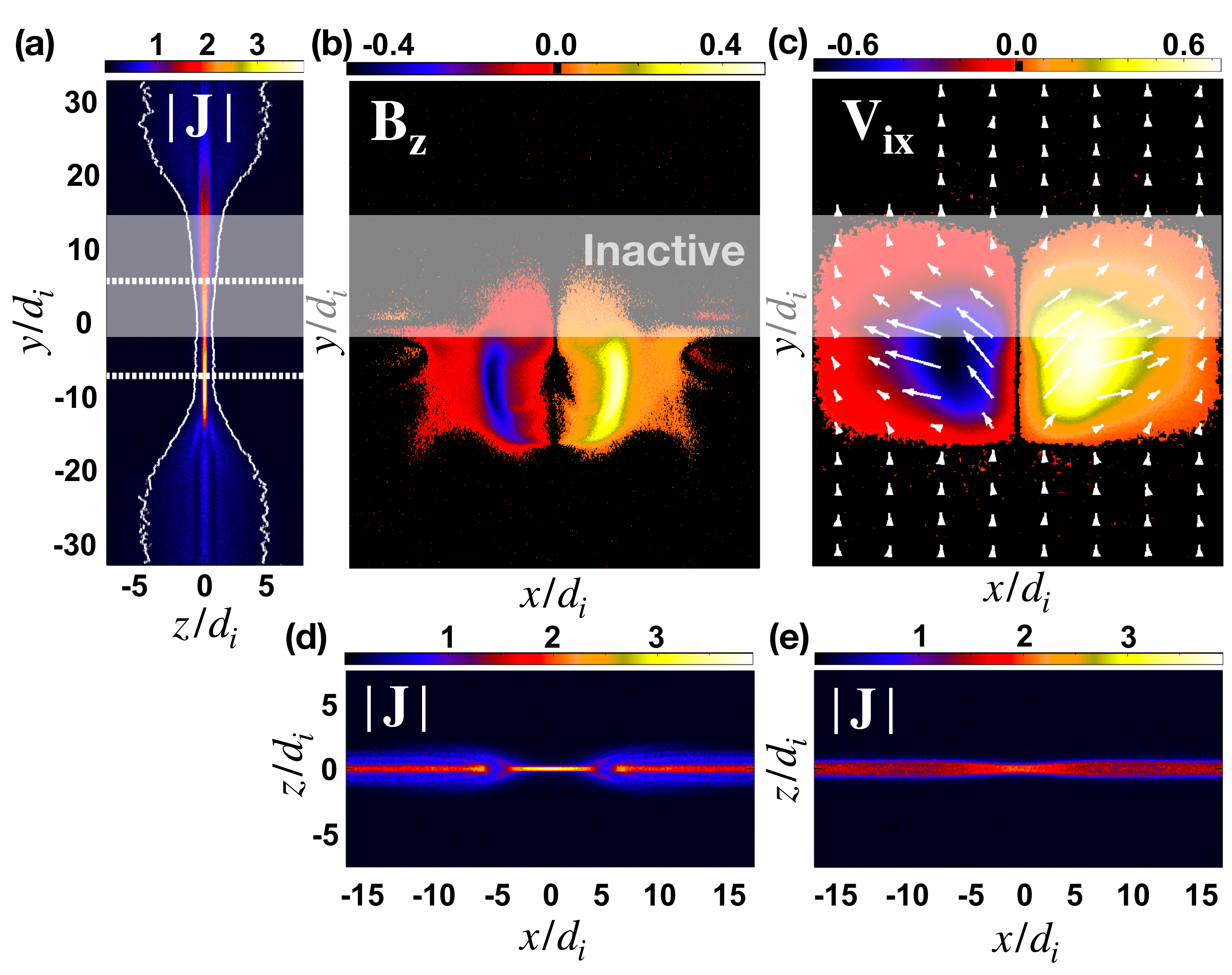} 
\caption {The 3D structure of reconnection with $L_{y,thin}$$=31d_i$ at time $12/\Omega_{ci}$. In (a) the current density $|{\bf J}|$ at the $x=0$ plane. In (b) the reconnected field $B_z$ and in (c) the ion outflow speed and the ${\bf V}_{i}$ vectors in white at the $z=0$ plane. In (d) and (e), the current density $|{\bf J}|$ on the $x-z$ plane along the lower and upper dashed lines in (a), respectively.
}
\label{L20_3D}
\end{figure}

In Fig.~\ref{L20_3D}, we look into the 3D structure of the reconnection region at time $12/\Omega_{ci}$, after the reconnection rate reaching its maximum (i.e., check Fig.~\ref{rates}). For reference, the current density at the $x=0$ plane is shown again in Fig.~\ref{L20_3D}(a). The reconnected field $B_z$ at the $z=0$ plane is shown in panel (b), and the ion outflow speed $V_{ix}$ is shown in panel (c). Black regions cover the region of zero value, contrast the reconnecting region of colors. The x-line extent is revealed between the region of opposite $B_z$ polarity near $x=0$, and the true extent can still be approximated by $L_{y,thin}=31d_i$. One pronounced feature is the asymmetric distribution of reconnection signatures in the y-direction. The $B_z$ signature is clearly shifted to the electron-drifting side. Inside this $d_i$-scale thin current sheet, it consists of two regions; one is the active region on the electron-drifting side with strong $B_z$ and $V_{ix}$ signatures. Another region on the ion-drifting side has weaker $B_z$ and $V_{ix}$, indicating a {\it weaker} reconnection; we refer it as the ``inactive'' region hereafter and we mark it with transparent white (or yellow) bands. Note that the extent of this inactive region is around $\simeq 10d_i$. In panel (d), we make a $x-z$ slice of the current density $|{\bf J}|$ at the active region (along the lower horizontal white dashed line indicated in panel (a)). The morphology of the reconnection region is similar to that of a corresponding 2D simulation (not shown). For comparison, in panel (e) we make a similar slice at the inactive region (along the upper white horizontal dashed line indicated in panel (a)). The current sheet near the x-line is thicker in comparison to that of the active region in panel (d).

Here we would like to point out that this two-region scenario is similar to that observed in the two-fluid simulations \cite{meyer15a}. However, the ``inactive region'' in PIC simulations has a localized x-line geometry on the $x-z$ plane, while the ``inactive region'' in two-fluid model is more like a Sweet-Parker reconnection that has a long extended current sheet. The difference between two-fluid and kinetic descriptions of this region is interesting, indicating that the nature of the dissipation process plays a significant role in the results. \\

{\it \bf 5. $L_{y,thin}=8.4d_i$ case--}
Here we show what happened if the extent of the thin current sheet is comparable or smaller than the extent of this inactive region discovered in the previous section. As already shown in Fig.~\ref{rates}, both the reconnection rate and outflow speed drop significantly when $L_{y,thin}\lesssim 10d_i$, suggesting a switch-off of reconnection. Here we look into the details of the current sheet structure of the $L_{y,thin}=8.4d_i$ case and describe the general property of having $L_{y,thin} \lesssim 10d_i$. 

\begin{figure}
\includegraphics[width=8.0cm]{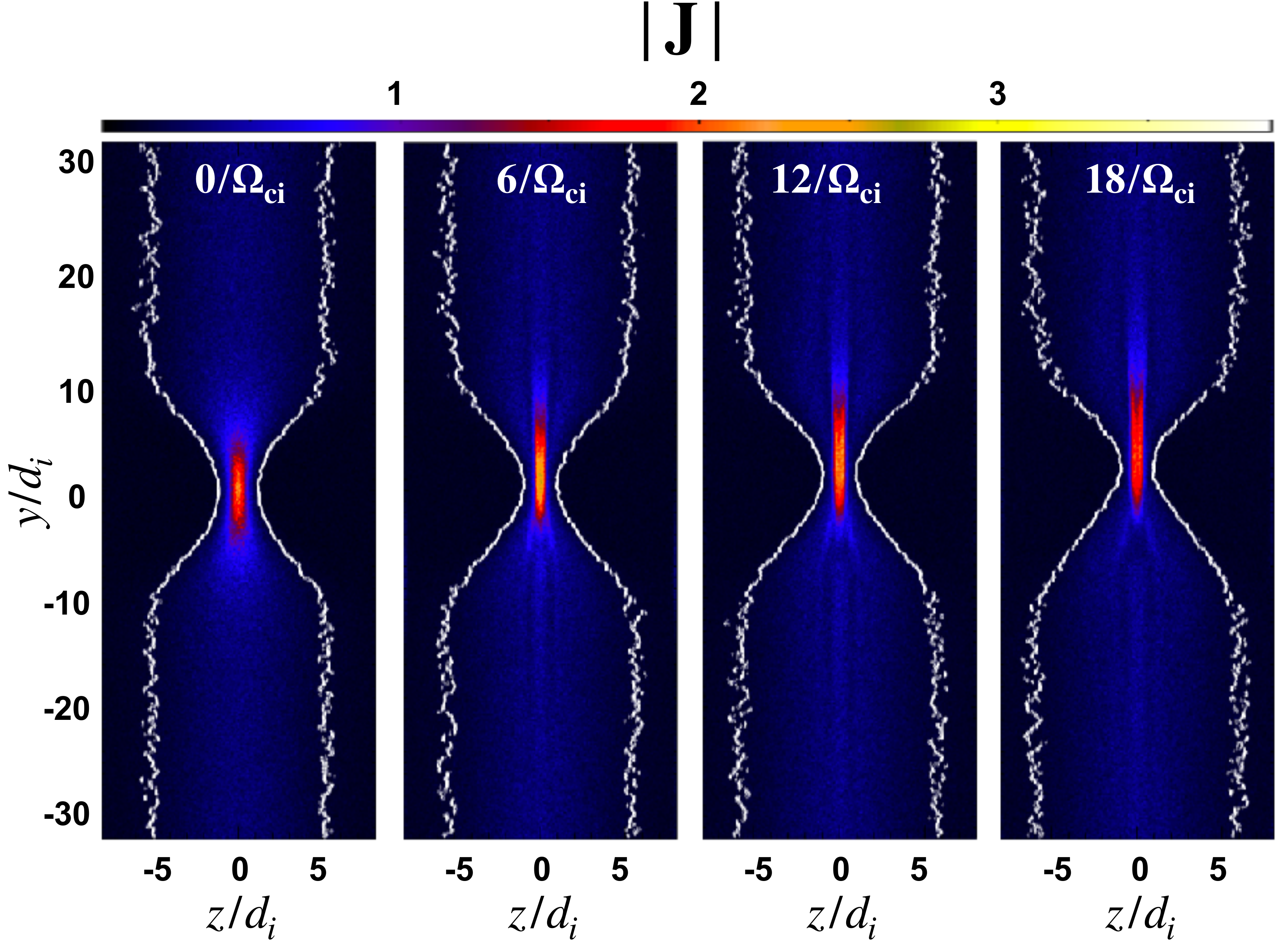} 
\caption {The evolution of the current density $|{\bf J}|$ at the $x=0$ plane inside the 3D box with $L_{y,thin}$$=8.4d_i$. The white curves trace the boundary of the current sheet.}
\label{L7d5_evolve}
\end{figure}

The evolution of the total current density $|{\bf J}|$ of the $L_{y,thin}=8.4d_i$ case at the $x=0$ plane (right through the x-line) is shown in Fig.~\ref{L7d5_evolve}. The corresponding times are $0/\Omega_{ci}$, $6/\Omega_{ci}$, $12/\Omega_{ci}$, and $18/\Omega_{ci}$, the same as that discussed for the $L_{y,thin}=31d_i$ case. The asymmetric thinning of the current sheet along the x-line is still recognizable, but the thinnest sheet on the electron-drifting side is not as thin as that at the active region of the $L_{y,thin}=31 d_i$ case shown in Fig.~\ref{L20_evolve}. As a result, this case does not reach fast reconnection locally on the electron-drifting side and reconnection is strongly suppressed. We will discuss how this asymmetric thinning connects to the reconnection process in the next section.

\begin{figure}
\includegraphics[width=9.0cm]{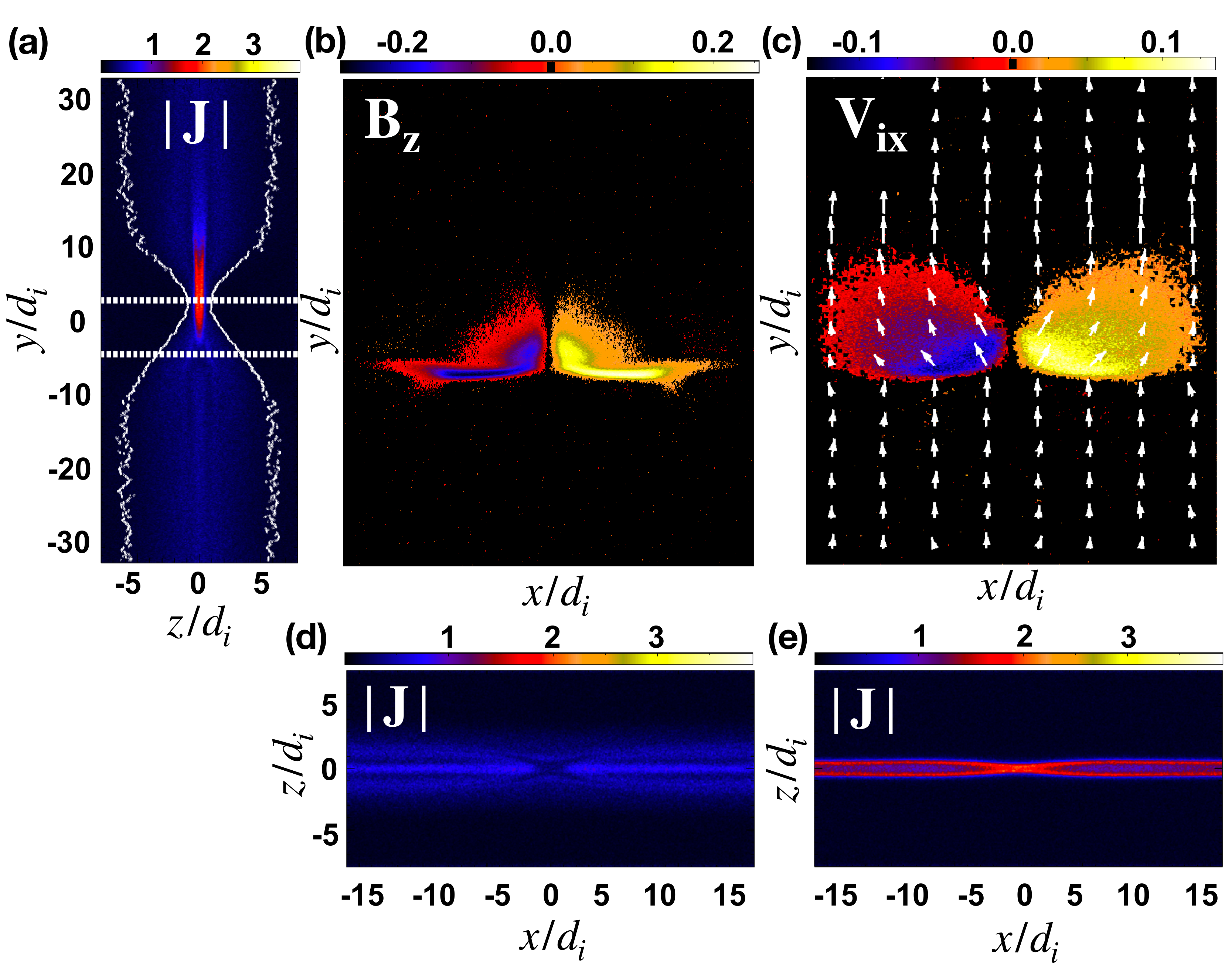} 
\caption {The 3D structure of reconnection with $L_{y,thin}$$=8.4d_i$ at time $18/\Omega_{ci}$. In (a) the current density $|{\bf J}|$ at the $x=0$ plane. In (b) the reconnected field $B_z$ and in (c) the ion outflow speed and the ${\bf V}_{i}$ vectors in white at the $z=0$ plane. In (d) and (e), the current density $|{\bf J}|$ on the $x-z$ plane along the lower and upper dashed lines in (a), respectively.
}
\label{L7d5_3D}
\end{figure}

The format in Fig.~\ref{L7d5_3D} is the same as that in Fig.~\ref{L20_3D}. The ion outflow speed $V_{ix}$ (in panel (c)) is reduced by $\simeq 6$ times compared to that in Fig.~\ref{L20_3D}. It becomes clear that both the reconnected field $B_z$ (in panel (b)) and the outflow speed $V_{ix}$ become narrower in y and concentrate on the electron-drifting side when $L_{y,thin}$ is smaller. Surprisingly, by comparing with panel (a), we realize that part of these more intense signatures are within the thick current sheet region. The real x-line extent manifested by the finite $B_z$ on the x-y plane can still be approximated as $L_{y,thin}=8.4 d_i$. Panel (d) shows the current sheet structure on the slice along the lower horizontal line in panel (a), that passes through the strong $B_z$ and $V_{ix}$ region. The current sheet is much thicker and the current density is reduced near $(x,z)=(0,0)$. As will be discussed in the next section, the reconnected field $B_z$ is swept into the thick current sheet but the magnetic tension $({\bf B}\cdot \nabla){\bf B}/4\pi \simeq B_z\partial_zB_x/4\pi$ associated with the reconnected field lines remains active in driving outflows, although with a reduced speed. \\

{\it \bf 6. The extent of the inactive region--}
The comparison of these two cases suggests the importance of the scale of this inactive region that fully develops within a long $L_{y,thin}$ current sheet. When $L_{y,thin} < L_{y,inact}\simeq O (10 d_i)$, it appears that the current sheet can not thin toward the thickness required for fast reconnection, and thus reconnection is strongly suppressed. The extent of this inactive region persists to have a similar y-extent at later time as indicated in the structure of the reconnected magnetic field $B_z$ in Fig.~\ref{tearing}. Also note that, at a later time $t=18/\Omega_{ci}$ a secondary tearing mode is generated on the electron-drifting side, which further maps out the thinnest region of the entire x-line. An important question is then how to determine the spatial scale of this inactive region. \\

\begin{figure}
\includegraphics[width=9.0cm]{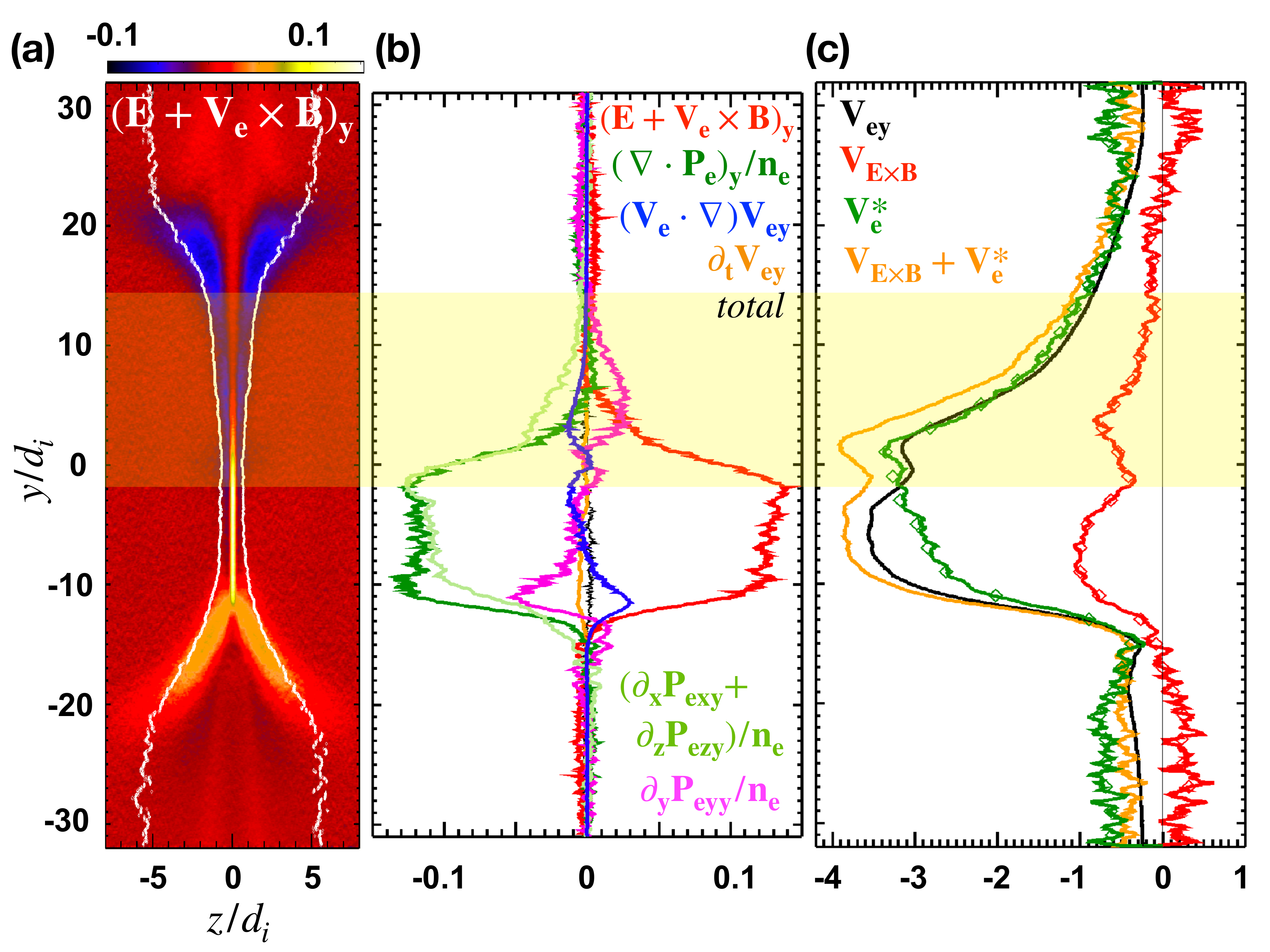} 
\caption {Analyses of the $L_{y,thin}=31d_i$ case at time $12/\omega_{ci}$. Panel (a) shows the non-ideal electric field $({\bf E}+{\bf V_e}\times {\bf B}/c)_y$ at the $x=0$ plane. Panel (b) shows the decomposition of the non-ideal electric field along the $(x,z)=(0,0)$ line. Panel (c) shows the decomposition of the electron drift near the x-line.}
\label{Ohms}
\end{figure}

{\it \bf 6. 1. 3D Ohm's law--}
To achieve fast reconnection in collisionless plasmas, the $d_i$-scale thin current sheet needs to thin further toward electron scale so that the frozen-in condition between electrons and magnetic fields can be broken. 
We quantify this effect using the generalized Ohms's law, which is basically the electron momentum equation,
\begin{equation}
{\bf E}+\frac{{\bf V}_e\times {\bf B}}{c}=-\frac{\nabla\cdot{\bf P}_e}{e n_e}-\frac{m_e}{e}{\bf V}_e\cdot\nabla {\bf V}_e-\frac{m_e}{e}\partial_t {\bf V}_e. 
\label{Ohms_law}
\end{equation}
The left-handed side measures the non-ideal electric field that is supported by the non-ideal terms on the right-handed side.
The y-component of the non-ideal electric field is relevant to the reconnection electric field and its structure at the $x=0$ plane is shown in Fig.~\ref{Ohms}(a). Within the active region between $y\in [-12,-2]d_i$, the magnitude of the non-ideal electric field $E_y+({\bf V}_e\times {\bf B})_y/c$ is $\simeq 0.12 B_x V_{Ax}$, consistent with the typical value of the fast reconnection rate \cite{yhliu17a,cassak17a}. The contributions of the non-ideal terms along the x-line are plotted in Fig.~\ref{Ohms}(b). Note that the ``{\it total}'' in black color sums up all terms and is negligible, indicating the excellent accuracy of this calculation. The $\partial_t V_{ey}$ term in orange color is also negligible, indicating a rather quasi-steady state.  
Consistent with the standard 2D simulation, the non-ideal electric field in the active region is supported by the divergence of the pressure tensor $\nabla\cdot {\bf P}_e$ of which the primary contribution comes from the off-diagonal component, $\partial_x P_{exy}+\partial_z P_{ezy}$. To filter out a potential contribution from an electrostatic component (instead of the electromagnetic component) that does not contribute to reconnection, we apply the General Magnetic Reconnection (GMR) theory \cite{hesse88a, schindler88a} to calculate the global 3D reconnection rate. To evaluate the global rate, it requires to integrate $E_\|$ along the magnetic field line that thread the ideal region to the localized non-ideal region, then back to the ideal region on the other side. Since we do not expect a significant difference if an infinitesimal guide field is applied, we will integrate $E_y$ along the x-line and note that $\int_0^{L_y} dy=\oint dy$ because of the periodic boundary condition in the y-direction. The generation rate of the total reconnected flux is $\oint E_y dy=2.1 B_x V_{Ax} d_i$, and the corresponding 2D rate is $(\oint E_y dy)/L_{y,thin}\simeq 2.1/31=0.068$, showing an excellent agreement with the value measured using $(\int{B_zdxdy})/L_{y,thin}$ in Fig.~\ref{rates}(a).

In contrast to a 2D model, now the $\partial_y$ terms survive in the 3D system. One of the new terms is $\partial_yP_{eyy}$ in $\nabla\cdot {\bf P}_e$, another is the electron inertia term ${\bf V}_e\cdot \nabla {\bf V}_{ey}=V_{ey}\partial _y V_{ey}$; note that both $V_{ex}$ and $V_{ez}$ vanish along the x-line due to the symmetry that coincides the flow stagnation point with the x-line.
The closed integration $\oint {V_{ey} \partial_y V_{ey}dy}=\oint {(1/2) \partial_y V_{ey}^2}dy=0$ and here $\oint {(1/en_e)\partial_yP_{eyy}dy}\simeq-0.018 B_x V_{Ax} d_i$ that is two-order smaller compared to the contribution from the off-diagonal contribution $\oint (1/en_e)(\partial_x P_{exy}+\partial_z P_{ezy}) dy$. These two terms thus do not contribute to the integral $\oint E_y dy$ in this 3D system, but they may re-distribute $E_y$. The term $\partial_yP_{eyy}$ contributes negatively to the non-ideal electric field on the ion-drifting side, positively on the electron-drifting side. One may argue that, perhaps, $\partial_yP_{eyy}$ on the ion-drifting side suppresses the typical fast reconnection electric field of order $0.1 B_x V_{Ax}$~\cite{yhliu17a,cassak17a}. Thus, balancing $0.1 B_x V_{Ax}\simeq (1/en_e)\partial_yP_{eyy}\simeq  (1/en_e)(B_x^2/8\pi)/L_{y,inact}$ could lead to a gradient scale $L_{y,inact}$ of an order $10d_i$ for the inactive region. (i.e., in the last step, one may argue that the pressure difference is $\Delta P\simeq B_x^2/8\pi$). However, the $\partial_yP_{eyy}$ term shown here as the pink curve of Fig.~\ref{Ohms}(b) is too small (compared to 0.1) to validate this argument. The electron inertia term $V_{ey}\partial _y V_{ey}$ contributes positively to the non-ideal electric field on the ion-drifting side, negatively on the electron-drifting side. Similarly, one may construct an argument to infer the gradient scale of this term by balancing it with the fast reconnection rate, but its magnitude as shown by the blue curve of Fig.~\ref{Ohms}(b) is also too small to be a valid explanation. \\

{\it \bf 6. 2. Time-scale toward fast reconnection and electron drifts--}
In 2D steady symmetric reconnection, the only non-ideal term that can break the frozen-in condition right at the x-line is the divergence of the off-diagonal component of the pressure tensor, $\partial_x P_{exy}+\partial_z P_{ezy}$. For this term to be significant, it requires the current sheet to be thin enough and comparable to the electron gyro-radius scale ($\rho_e$) so that the nongyrotropic feature develops \cite{hesse11a}. (Here $\rho_e \simeq 0.61d_e=0.07d_i$ based on the initial electron pressure at the thin sheet and the reconnecting field). Thus, to reach fast reconnection the current sheet thinning is an unavoidable route. The tension force $B_z\partial_zB_x/4\pi$ rising from the reconnected magnetic flux $B_z$ is required to drive outflow, that leads to current sheet thinning. In a 3D system, we have an additional transport of this normal flux ($B_z$) in the electron drift direction below the $d_i$-scale; because ions are de-magnetized while electrons are still magnetized (i.e., the Hall effect). This transport removes this flux from what becomes the inactive part of the x-line. This removal of $B_z$ prevents outflows, and, hence, thinning of the current sheet. As a consequence, the current sheet thickness in the $L_{y,thin}=8.4d_i$ case can not reach the thinnest thickness as that in the $L_{y,thin}=31d_i$ case, and this appears to throttle reconnection.

The electron drift speed along the anti-current (-y) direction consists of the ${\bf E}\times{\bf B}$ drift and the diamagnetic drift, ${\bf V}_{e,\perp}\approx c({\bf E}\times{\bf B})_y/B^2-c({\bf B}\times \nabla\cdot {\bf P})_y/en_eB^2$. The primary components are
\begin{equation}
V_{ey}\approx c\frac{E_zB_x}{B^2}+\frac{c B_x \partial_z P_{ezz}}{en_e B^2}.
\label{Vey}
\end{equation}
The diamagnetic drift (${\bf V}^*$ in green) dominates the electron drift within this thin current sheet, as shown in Fig.~\ref{Ohms}(c). Note that a diamagnetic drift can also transport the magnetic flux even though the guiding centers of electrons do not really move (i.e., roughly speaking, we can swap $x$ and $y$, and assume $B_z \ll B_x$ in Eq.(\ref{Vey}) here to recover Eq.(1) of Liu and Hesse \cite{yhliu16a} that transports the reconnected flux as indicated in Eq.(2) therein. See also \cite{swisdak03a,coppi65a}).
This preferential flux-transportation by electrons results in the enhanced reconnected magnetic flux $B_z$ on the electron-drifting sides shown in Fig.~\ref{L20_3D}(b) and \ref{L7d5_3D}(b). This transport also explains why the current sheet only becomes thinner on the electron-drifting side as shown in Fig.~\ref{L20_evolve} and \ref{L7d5_evolve}, and the preferential occurrence of the secondary tearing mode on the electron-drifting side as shown in Fig.~\ref{tearing}.

One can then imagine that the time-scale of the current sheet thinning process toward fast reconnection can be translated into the spatial-scale of the inactive region, and it is
\begin{equation}
L_{y,inact}\simeq T_{thinning}\times V_{ey}. 
\label{length}
\end{equation}
The electron drift speed is on the order of $V_{Ax}$ inside this inactive region. On the other hand, reconnection in the $L_{y,thin}=31d_i$ case reaches the maximum rate at time $\simeq 10/\Omega_{ci}$ as shown by the blue curve in Fig.~\ref{rates}(a), thus $T_{thinning}\simeq 10/\Omega_{ci}$. (Note that this time-scale in 3D is comparable to the time-scale of the companion 2D simulation shown in black color). The rough estimation of Eq. (\ref{length}) suggests that the extent of this inactive region should be on the order of $L_{y,inact}\simeq 10/\Omega_{ci} \times V_{Ax} = 10d_i$, which agrees with the observed spatial-scale. More accurately, we can integrate the time for the flux to be transported within the inactive region (marked by the yellow band that spans $y\in [-2, 14]d_i$) using the $V_{ey}$ profile in Fig.~\ref{Ohms}(c). It estimates the transport time-scale $T_{transport}=\int (dy/V_{ey})\simeq 10/\Omega_{ci}$ that compares favorably to the thinning time-scale $T_{thinning}$ just discussed. This quantitative examination validates this flux-transport mechanism in determining the extent of the inactive region. \\

{\it \bf 7. Summary and discussion on the dawn-dusk asymmetry--}
We modified the Harris sheet geometry to embed an inertial-scale ($d_i$) thin current sheet between much thicker sheets in the current direction. The resulting reconnection is well confined within the thin current sheet. With this machinery, we investigate the shortest possible x-line extent for fast reconnection, which appears to be $\simeq 10d_i$. The time-scale for a $d_i$-scale current sheet to thin toward the condition suitable for fast reconnection (with the normalized reconnection rate $\simeq 0.1$) can be translated into an intrinsic length-scale $\simeq 10d_i$ of an inactive region after considering the flux transport along the x-line (Eq.(\ref{length})); because the reconnected magnetic flux ($B_z$) required to drive outflows and further the current sheet thinning is transported away in the anti-current direction by electrons below the ion inertial scale (i.e., the Hall effect). 
We do not expect a strong dependence of this critical length on the mass ratio. The nonlinear growth time of reconnection appears to be virtually independent on mass ratio, and so does the flux transport; this is consistent with the apparent independence of the reconnection rate on the mass ratio \cite{shay98a,hesse99a}.
Simulations demonstrate that reconnection is strongly suppressed if the extent of the thin current sheet is shorter than this intrinsic length-scale of the inactive region. In these short $L_{y,thin}$ cases, the outflow driver $B_z$ is completely removed from the reconnecting region. The current sheet thus is not able to thin to the thickness where the nongyrotropic feature of the electron pressure tensor develops and becomes significant for breaking the {\it frozen-in} condition at the x-line. 

\begin{figure}
\includegraphics[width=8.0cm]{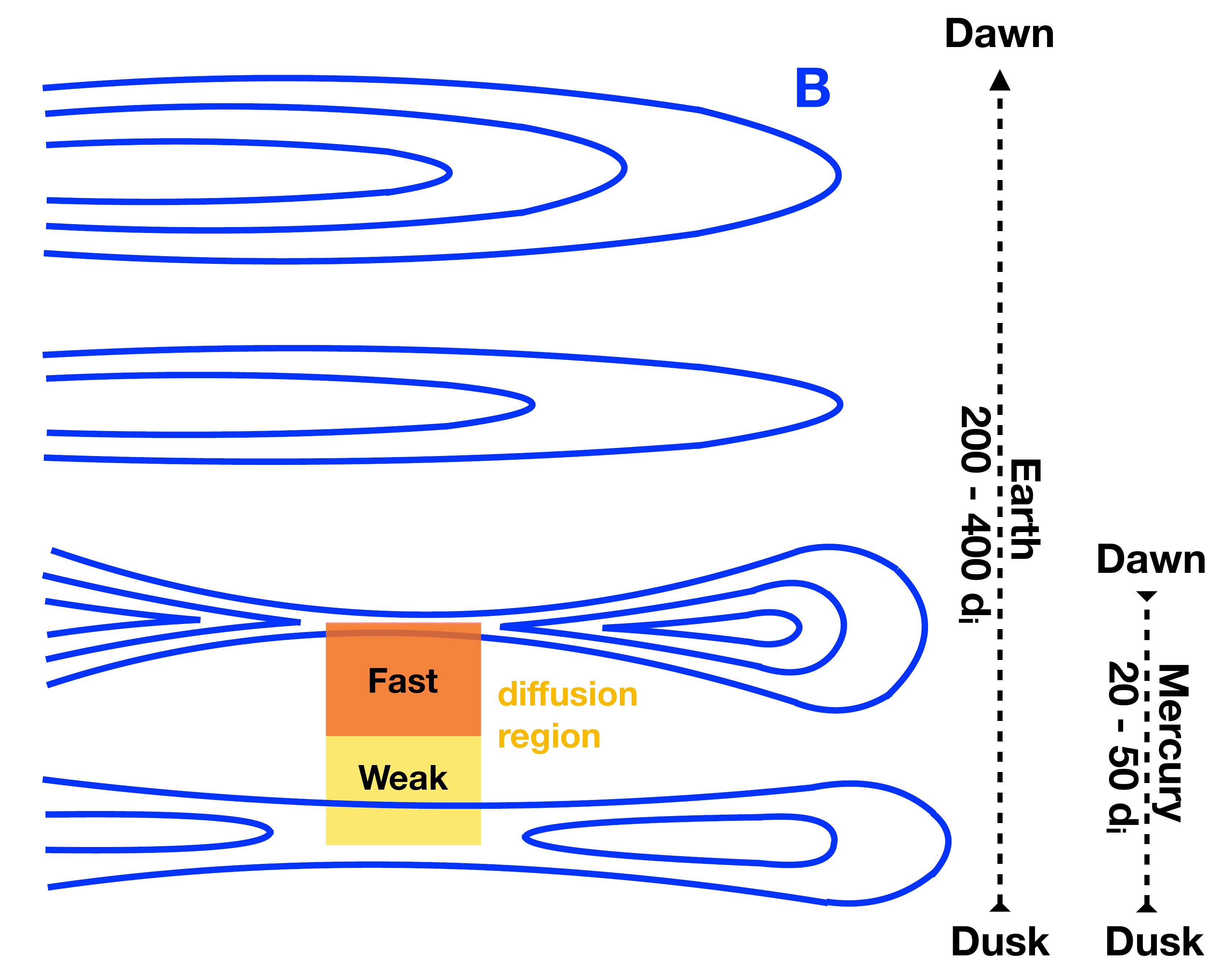} 
\caption {An explanation of why the dawn-dusk asymmetry is opposite at Earth and Mercury based on the dawn-ward transport of normal magnetic fields ($B_z$) and reconnection physics. (Note that in this figure the dawn and dusk sides are switched vertically to follow the convention).
}
\label{dawn_dusk}
\end{figure}

Reconnection is strongly suppressed when the x-line extent is shorter than the length-scale of the inactive region $L_{y,inact}\simeq O(10 d_i)$, and this may explain the narrowest possible dipolarizing flux bundle (DFB) observed at Earth's magnetotail \cite{JLiu15b}. Note that an interchange/ballooning instability may locally trigger reconnection [e.g.,\cite{pritchett13a}] and our basic conclusion on the minimal x-line extent should still hold in the complex coupling to an instability.  On the other hand, this internal dawn-dusk asymmetry of the reconnection x-line (e.g., Fig~\ref{tearing}) may also explain why the flux transport events occur preferentially on the dawn side of Mercury's magnetotail \cite{WSun16a}. The fact that the active region preferentially occurs on the electron-drifting side (i.e., the dawn side) seems to contradict to the explanation of the dawn-dusk asymmetry discussed in Lu et al. \cite{SLu16a, SLu18a}. Here we clarify the similarity and difference of our studies, which leads to a plausible explanation to the opposite dawn-dusk asymmetry observed at Earth \cite{slavin05a,nagai13a,runov17a} and Mercury \cite{WSun16a}. While the electron drift transports the normal magnetic flux ($B_z$) in both studies, the important difference stems from the role of the normal magnetic field ($B_z$) discussed. In Lu et al. \cite{SLu16a, SLu18a}, the initial normal magnetic field $B_z$ associated with the tail geometry suppresses the onset of reconnection since it prevents the current sheet from being tearing unstable \cite{hesse01a,yhliu14b,sitnov10a}. Reconnection onsets are thus easier on the dusk side since these $B_z$ flux is transported to the dawn side. In contrast, the reconnected field ($B_z$) discussed here drives outflows and furthers the thinning toward fast reconnection after reconnection onset. 
As illustrated using Fig.~\ref{dawn_dusk}, the explanation of the dawn-dusk asymmetry in Lu et al. \cite{SLu16a, SLu18a} can remain valid in predicting the global asymmetry of reconnection ``onset locations'' on the dusk side of Earth. While our study explains the ``internal'' asymmetric structure of the x-line within these onset locations, that gives rise to the active region on the dawn side locally. 

For Mercury, if one considered a proton density of $\sim 3 cm^{-3}$ \cite{gershman14a,WSun18a,poh18a}, and the relatively thin current sheet width in Mercury's tail near midnight is $\sim 2 R_M$ where $1R_M \sim 2440$ km \cite{WSun16a, poh17b, rong18a}, then the global dawn-dusk extent is $\sim 37 d_i$, comparable to our $31d_i$ case studied here. While for Earth, the proton density in the plasma sheet is around an order of magnitude smaller than that at Mercury \cite{baumjohann89a, CHuang94a, WSun18a}, and the width of the relatively thin current sheet near midnight is $\sim 20 R_E$ \cite{nakai91a, SZhang16a}, corresponding to $\sim 300 d_i$. The dawn-dusk extent of the thin current sheet region at the magnetotail of Mercury is thus much shorter (in terms of $d_i$) than that of Earth. Therefore, the entire magnetotail of Mercury likely only manifests the internal dawn-dusk asymmetry of the x-line with the active region and secondary tearing modes appearing on the dawn side, as emphasized by the orange region of Fig.~\ref{dawn_dusk}. We further predict that magnetic reconnection may not occur in a planetary magnetotail if its global dawn-dusk extent is $\ll 10d_i$. Finally, while these arguments are purely based on the reconnection physics in the plasma sheet, we acknowledge that global effects [e.g., \cite{walsh14a, lotko14a, spence93a, keesee11a}] could also be important but are beyond the scope of this study. \\

\acknowledgments 
Y.-H. Liu thanks S. Lu, S. Wang, A. M. Keesee and M. Shay for helpful discussions. 
YHL and TCL were supported by NASA grant 80NSSC18K0754 and MMS mission. MH was supported by the Research Council of Norway/CoE under contract 223252/F50, and by NASA's MMS mission. JL was supported by NSF grant 1401822 and NASA contract NAS5-02099. JAS was supported by NASA MMS GI grant 80NSSC18K1363. Simulations were performed with NERSC Advanced Supercomputing, LANL institutional computing and NASA Advanced Supercomputing. The large data set generated by 3D PIC simulations can hardly be made publicly available. Interested researchers are encouraged to contact the leading author for subsets of the data archived in computational centers.\\


\end{document}